\documentclass[pr,twocolumn,
showkeys,showpacs,nofootinbib,byrevtex]{revtex4}%{revtex4-1}
\usepackage{graphicx,amsmath,amsfonts}
\usepackage{amssymb,amscd,amsxtra,amsthm}
\usepackage{epsfig,bm,dcolumn}
\usepackage{epstopdf}
\usepackage{xcolor}

\begin{document}

\title{Naive parton picture for kaon color transparency in $A(e,e'K^+)$}

\author{Kook-Jin Kong}
\thanks{kong@kau.ac.kr}
\affiliation{Research Institute of Basic Science, Korea Aerospace University, Goyang 10540, Korea}

\author{Tae Keun Choi}
\thanks{tkchoi@yonsei.ac.kr}
\affiliation{Department of Physics and Engineering Physics, Yonsei University, Wonju 26493, Korea}

\author{Byung-Geel Yu}
\thanks{bgyu@kau.ac.kr}
\affiliation{Center for Exotic Nuclear Studies, Institute for
Basic Science, Daejeon 34126, Korea}
\affiliation{Research
Institute of Basic Science, Korea Aerospace University, Goyang
10540, Korea}

%%%%%%%%%%%%%%%%%%%% Abstract %%%%%%%%%%%%%%%%%%%%%

\begin{abstract}
Nuclear transparency in the electronuclear reaction $A(e,e'K^+)$ is investigated
in parallel with our previous study of pion transparency in
Phys.\ Rev.\ C {\bf 111}, 064608 (2025). Based on an extended Glauber
framework that incorporates shadowing from the initial-state two-step
process, kaon color transparency (CT) is analyzed to show that the
steeper $Q^2$ dependence observed for kaon CT, compared with the pion
case, is more naturally described by the naive parton model (NPM) than
by the quantum diffusion model (QDM). The inclusion of initial-state
shadowing further reduces the transparency and improves the agreement
with the experimental data. The $Q^2$ and $A$ dependences of the kaon
transparency are presented up to $Q^2=10$~GeV$^2/c^2$, together with the
corresponding $\alpha(Q^2)$ and the supplementary ratio $T_A/T_C$, for
comparison with the Jefferson Lab (JLab) data obtained with the 6-GeV
electron beam on $^{12}$C, $^{63}$Cu, and $^{197}$Au nuclei.
\end{abstract}



\maketitle

\section{Introduction}

Nuclear transparency is a particularly useful observable for
studying hadron propagation in nuclei and, at sufficiently large
momentum transfer, the onset of color transparency (CT) in
QCD~\cite{Frankfurt1994,Dutta2013}. In the conventional hadronic
picture, the attenuation of the ejectile is governed by its
interactions with the nuclear medium, whereas CT is associated
with the production of a small-size color-singlet configuration
whose interaction with surrounding nucleons is reduced during the
early stage of its expansion~\cite{Frankfurt1994,Dutta2013}. Meson
electroproduction is especially attractive in this respect, since
CT is generally expected to set in more favorably for $q\bar q$
systems than for baryons, and the JLab pion-transparency
measurements indeed revealed a systematic rise of the transparency
with $Q^2$ over a broad range of
nuclei~\cite{Clasie2007,Qian2010,Dutta2013}.

Kaon electroproduction on nuclei is of particular interest in this
context. Compared with the pion sector, dedicated studies of kaon
transparency remain rather limited, even though the $K^+$ channel
offers a natural opportunity to test CT in the strangeness
sector~\cite{Nuruzzaman2011,Das2019,Dutta2013}. Experimentally,
Nuruzzaman \textit{et al.} extracted nuclear transparencies from
$A(e,e'K^+)$ on $^{12}\mathrm{C}$, $^{63}\mathrm{Cu}$, and
$^{197}\mathrm{Au}$ at $Q^2=1.1$, $2.2$, and
$3.0~\mathrm{GeV}^2$~\cite{Nuruzzaman2011}.
The extracted effective in-medium kaon-nucleon cross sections were
found to be smaller than the corresponding free-space values
inferred from simple geometrical attenuation
models~\cite{Nuruzzaman2011}. On the theory side, Das showed that
a conventional treatment based on free $K^+N$ cross sections
underestimates the measured transparency, whereas the introduction
of a distance-dependent effective kaon--nucleon cross section,
simulating CT, leads to a substantially improved
description~\cite{Das2019}.

The present work is motivated by the observation that the standard
Glauber framework, widely used in semiclassical analyses of meson
transparency and CT~\cite{Frankfurt1994,Larson2006}, need not
exhaust all relevant reaction mechanisms in electroproduction on
nuclei. In our previous study of pion transparency, we went beyond
the conventional Glauber treatment by incorporating the shadowing
generated by the initial-state two-step process associated with
the $\gamma^\ast\to\rho^0$ fluctuation~\cite{Choi2025}. We found
that, although the usual Glauber treatment supplemented with CT
reproduces the gross $Q^2$ dependence, it tends to overestimate
the measured pion transparency, whereas the additional
vector-meson-induced shadowing leads to a substantially better
description of the JLab $A(e,e'\pi^+)$ data for $^{12}\mathrm{C}$,
$^{27}\mathrm{Al}$, $^{63}\mathrm{Cu}$, and
$^{197}\mathrm{Au}$~\cite{Choi2025}. This naturally motivates a
corresponding reexamination of kaon transparency, where the
interplay of final-state CT and initial-state shadowing may also
be relevant.

More microscopic descriptions of nuclear attenuation and CT
certainly exist, including the Green-function/path-integral
approach of Kopeliovich and collaborators~\cite{Kopeliovich1993},
the relativistic multiple-scattering Glauber framework of the
Ghent group~\cite{Cosyn2008}, the semiclassical treatments of
Larionov and collaborators~\cite{Larionov2020}, and
coupled-channel transport calculations in
GiBUU~\cite{Gallmeister2022}. While these approaches are valuable
for a more dynamical treatment of the reaction mechanism, their
practical implementation generally involves substantially greater
complexity and additional model dependence associated with
intermediate-state propagation, multichannel couplings, and
reaction-specific
ingredients~\cite{Kopeliovich1993,Cosyn2008,Larionov2020,Gallmeister2022}.
For the present purpose, namely to identify the dominant mechanisms
governing kaon transparency in a form that can be confronted directly
with the existing data, a transparent phenomenological framework is
particularly useful. In such dynamical approaches, Fermi motion and
Pauli blocking can influence the quantitative extraction of
attenuation, although their roles are different: the former mainly
smears the initial quasifree kinematics, whereas the latter restricts
the available baryonic final states~\cite{Qian2010,Nuruzzaman2011}.
In the present kaon channel, Pauli-blocking effects are expected to be
less direct than in pionic channels, since the produced $K^+$ is a
boson and the associated hyperon is not Pauli blocked by the nucleon
Fermi sea~\cite{Garibaldi2019}.

Motivated by these considerations, we revisit nuclear transparency
in $A(e,e'K^+)$ within an extended Glauber framework that includes
both final-state CT effects through a reduced effective
kaon--nucleon cross section and initial-state shadowing associated
with vector-meson fluctuations of the virtual photon. Our aim is
to determine how far the existing kaon-transparency data can be
understood from the interplay of these two mechanisms in a simple
phenomenological description, and to compare the kaon case with
our previous pion-transparency analysis in order to discuss the
onset of CT in the strangeness sector.

This paper is organized as follows. In Sec.~II, we formulate the
NPM~\cite{farrar} alongside the QDM for analyzing the $Q^2$
dependence of kaon transparency in the reaction $A(e,e'K^+)$,
based on the extended Glauber framework developed in
Ref.~\cite{Choi2025}. The shadowing effect introduced in
Ref.~\cite{Choi2025} is included as an additional mechanism that
further reduces the apparent kaon transparency. Within this
framework, we reproduce the observed $Q^2$ and $A$ dependences of
the nuclear transparency $T_A$ and evaluate the corresponding
$Q^2$ dependence of the parameter $\alpha$. As a supplementary
observable, we also present the ratio $T_A/T_C$, i.e.\ the
transparency normalized to that of $^{12}$C. These results are
then compared with the JLab data obtained with the 6-GeV electron
beam. Section~III contains a summary and concluding remarks.

\section{Models for data analysis}

The reaction $A(e,e'K^+)$ is induced by the elementary subprocess
$
\gamma^*(k)+p(p)\to K^+(p_K)\Lambda(p') \ (\text{or }
\Sigma^0),
$
with the relevant kinematical quantities listed in
Table~\ref{tab:kinematics}. In the present work, the nuclear
transparency $T_A$ is evaluated within an extended Glauber
framework that incorporates both the final-state attenuation of
the outgoing kaon and the initial-state shadowing associated with
the two-step process. The resulting expression is given by~\cite{Choi2025}
\begin{eqnarray}
&&T_A=\frac{1}{A} \int d^2b \int_{-\infty}^{\infty} dz\,
\varrho(b,z)\times
\nonumber\\%
&&\exp\left[ -\sigma_{\phi N}\int_{z-l_c}^{z} dz'\varrho(b,z')
-\int_{z}^{\infty} dz'\,\sigma^{\rm eff}_{KN}(z')\varrho^{*}(b,z')
\right].
\nonumber\\
\label{eq:TA}
\end{eqnarray}
Here $\varrho$ and $\varrho^{*}$ denote the nuclear density
distributions of the target nucleus $A$ and the recoiling $(A-1)$
system, respectively. In the present calculation we employ a
Woods--Saxon distribution for the nuclear density with the same
prescription as in Ref.~\cite{Choi2025}.

%%%%%%%%%%%%%%%%%%%%%%%%%%%%%%%%% TB %%%%%%%%%%%%%%%%%%%%%%%%%%%%%%%%%%%%%%%%%%%%%%%%%
\begin{table}[t]
\caption{ Kinematical variables $Q^2$ (GeV$^2/c^2$), $W$ (GeV),
$E_e$ (GeV), $\theta_e$ (deg), $E'_e$ (GeV), $x_B$, and $p_K$
(GeV$/c$) for kaon electroproduction, together with the quantities
relevant to the JLab measurement of kaon transparency in the range
$0.5\le Q^2\le 10$ GeV$^2/c^2$~\cite{Nuruzzaman2011}. Also shown
are the coherence length $l_c$ (fm) with $m_\phi$ in
Eq.~(\ref{eq:lc}), the formation length $l_f^{\rm NPM}$ (fm) with
$R_K=0.58$ fm in Eq.~(\ref{eq:lfnpm}), and the formation length
$l_f^{\rm QDM}$ (fm) with $\Delta M^2=0.3$ GeV$^2$ in
Eq.~(\ref{eq:lfqdm}).
For each kinematical setting,
the total cross section $\sigma_{KN}(p_K)$ is assigned from the PDG~\cite{pdg}
$K^+N$ total-cross-section values at the corresponding kaon momentum
$p_K$. Since the variation over the relevant momentum range is weak,
this amounts in practice to using piecewise-constant values, as listed
in the last column of the table.
Throughout this work we use
$\sigma_{\phi N}=18$ mb for the shadowing correction associated
with the two-step process~\cite{Choi2025}. }
\begin{ruledtabular}
\begin{tabular}{lllllllllll}
$Q^2$& $W$  & $E_{e}$&$\theta_e$& $E_e'$&$x_B$&$p_K$&$l_c$&$l^{\rm{NPM}}_f$&$l^{\rm{QDM}}_f$&$\sigma_{KN}$ \\
\hline
0.55 &2.28  &3.4  &26   &0.8 &0.11&2.24 &0.65 &2.7& 2.95& \\
0.69 &2.25  &3.5  &27   &0.9 &0.14&2.22 &0.60 &2.68& 2.93 &\\
0.77 &2.25  &3.68 &26   &1.04&0.16&2.26 &0.58 &2.71& 2.97 &18\\
0.92 &2.26  &3.85 &27   &1.1&0.18&2.36 &0.56  &2.83& 3.10& \\
1.05 &2.26  &4.02 &27   &1.2&0.2 &2.41 &0.54  &2.89& 3.17& \\
\hline
1.10 &2.26  &4.02 &27.76&1.19&0.21&2.42 &0.53&2.9& 3.18& \\
2.15 &2.21  &5.01 &28.85&1.73&0.35&2.74 &0.41&3.27&3.60&\\
3.0  &2.14  &5.01 &37.77&1.43&0.45&2.89 &0.35&3.45& 3.81& 17\\
3.91 &2.26  &5.77 &40.38&1.42&0.48&3.61 &0.35&4.29& 4.76& \\
4.69 &2.25  &5.77 &52.67&1.03&0.53&3.90 &0.33&4.62& 5.14& \\
\hline
5.0  &2.43  &11.0 &16.28&5.67&0.5 &4.57 &0.35&5.4& 6.00& \\
6.5  &2.74  &11.0 &22.13&4.01&0.5 &6.25 &0.37&7.37& 8.23& 18\\
8.0  &3.02  &11.0 &32.37&2.34&0.49&7.94 &0.38&9.34& 10.44& \\
9.5  &3.09  &11.0 &47.71&1.32&0.52&8.89 &0.36&10.46& 11.69& \\
\end{tabular}
\end{ruledtabular}
\label{tab:kinematics}
%\label{tb1}
\end{table}
%%%%%%%%%%%%%%%%%%%%%%%%%%%%%%%%%%%%%%%%%%%%%%%%%%%%%%%%%%%%%%%%%%%%%%%%%%%%%%%%%%%%%%%%%

Equation~(\ref{eq:TA}) makes the reaction sequence explicit. The
first exponential factor represents the shadowing correction in
the initial state, where the virtual photon fluctuates into a
vector meson at $z-l_c$ and propagates over the coherence length
before the kaon is produced at $z$ to scatter off a nearby nucleon~\cite{bauer,yennie,huefner}.
The second factor
describes the attenuation of the outgoing kaon through the
effective kaon--nucleon cross section $\sigma_{KN}^{\rm eff}(z)$
along its path through the nucleus. Since the virtual photon
couples to the kaon channel through the fluctuation $\gamma^*\to
\phi(1020)\to K^+K^-$, the coherence length is given by
\begin{equation}
l_c=\frac{2\nu}{Q^2+m_\phi^2}\,, \label{eq:lc}
\end{equation}
and the corresponding values are listed in
Table~\ref{tab:kinematics} in accordance with $Q^2$ values.

In the present framework, the elementary input for the hadronic
attenuation is crucial to determining the $Q^2$ dependence of the
transparency. In particular, the size of $\sigma_{KN}$ governs the
overall strength of the final-state interaction, while the choice
of $\sigma_{\phi N}$ controls the magnitude of the initial-state
shadowing. The cross section  $\sigma_{\phi N}$ is constrained
only indirectly through photoproduction data and
nuclear-transparency analyses, with typical free-space estimates
lying in the range of about $10$--$20$
mb~\cite{sibirt,Qian2009,das1,gubler}. In the present calculation,
we use the $K^+N$ total cross sections quoted by the
PDG~\cite{pdg}, as listed in Table~\ref{tab:kinematics}, and adopt
$\sigma_{\phi N}=18$ mb following Ref.~\cite{drell}.

To implement color transparency in the Glauber framework, one
needs a distance-dependent effective kaon--nucleon cross section.
A standard choice is the quantum diffusion model (QDM), whose
applicability has been examined extensively in the pion and
$\rho^0$ sectors~\cite{Dutta2013,Qian2010,larinov,Cosyn2008,aira,fassi}.
For the kaon case, however, the
observed $Q^2$ dependence appears steeper than what is naturally
generated by the usual QDM parametrization, even when the mass
parameter in the formation length is readjusted~\cite{Das2019}. This
motivates us to consider, in parallel, the naive parton model
(NPM), which offers a simple benchmark for the space-time
expansion of a compact $q\bar q$ configuration.

In the naive parton picture, the suppression of the effective
cross section is estimated from the reduced transverse size of the
initially produced configuration,
$
{\sigma_{KN}^{\rm eff}}/{\sigma_{KN}} \sim
{x_t^2}/{\langle x_t^2\rangle} \sim {\langle
k_t^2\rangle}/{Q^2},
$
where $x_t$ denotes the transverse size of the $q\bar q$ configuration.
Meanwhile,
partons separating at velocity $c$ propagate over the longitudinal
distance $z\sim (E/m)x_t$, which defines the formation length
after relativistic dilation~\cite{farrar}.
This gives
\begin{equation}
l_f^{\rm NPM}= \left(\frac{E_K}{m_K}\right)R_K\,,
\label{eq:lfnpm}
\end{equation}
where $R_K$ denotes the transverse size associated with the
forming kaon and is here taken to be the kaon charge radius. This
choice should not be interpreted as a strict identification of the
initially produced compact $q\bar q$ configuration with the
physical kaon, but rather as a practical estimate of the relevant
hadronic length scale without introducing an additional adjustable
parameter.
In contrast, the QDM associates the
expansion length with the inverse energy difference between
hadronic states~\cite{Dutta2013}, leading to
\begin{equation}
l_f^{\rm QDM}=\frac{2p_K}{\Delta M^2}\,. \label{eq:lfqdm}
\end{equation}

The effective cross section that accommodates both models can then
be written in the unified form
\begin{eqnarray}
&&\sigma_{KN}^{\rm eff}(z;Q^2,p_K)=\sigma_{KN}(p_K) \biggl[
\theta(z-l_f)
\nonumber\\%
&&+ \left\{ \frac{n^2\langle k_t^2\rangle}{Q^2}
\left(1-\left(\frac{z}{l_f}\right)^\tau\right) +
\left(\frac{z}{l_f}\right)^\tau \right\} \theta(l_f-z) \biggr],
\label{eq:sigmaeff}
\end{eqnarray}
where $\tau=2$ corresponds to the NPM and $\tau=1$ to the QDM, with
$n=2$ for the $q\bar{q}$ configuration. Equation~(\ref{eq:sigmaeff})
is the central ingredient of the present analysis, since it embeds the
CT effect directly into the nuclear absorption through the explicit
$Q^2$ dependence of $\sigma_{KN}^{\rm eff}$. In QCD, a virtual photon
with larger $Q^2$ selects a compact color-singlet $q\bar{q}$
configuration with a small transverse size, whose interaction with the
surrounding nucleons is suppressed by color screening. Since the
effective interaction strength decreases with the transverse size of
the compact configuration, the nuclear absorption becomes weaker as
$Q^2$ increases. Therefore, once ordinary nuclear effects such as
Glauber attenuation based on the free-space cross section and
shadowing are properly taken into account, a nontrivial $Q^2$
dependence of the effective absorption can be regarded as a direct
signature of the onset of color transparency~\cite{brodsky,jain,nikolaev}.

In the numerical calculation, we use $\langle
k_t^2\rangle^{1/2}\simeq 0.35$ GeV/$c$ for the quark transverse
momentum in the kaon. For each kinematical setting,
$\sigma_{KN}(p_K)$ is assigned from the PDG $K^+N$
total-cross-section values at the corresponding kaon momentum
$p_K$~\cite{pdg}. Because the variation over the relevant momentum
range is weak and no pronounced resonance structure appears there,
this is implemented in practice as piecewise-constant values, as
listed in Table~\ref{tab:kinematics}.

\begin{figure}[t]
\centering
\includegraphics[width=0.95\linewidth]{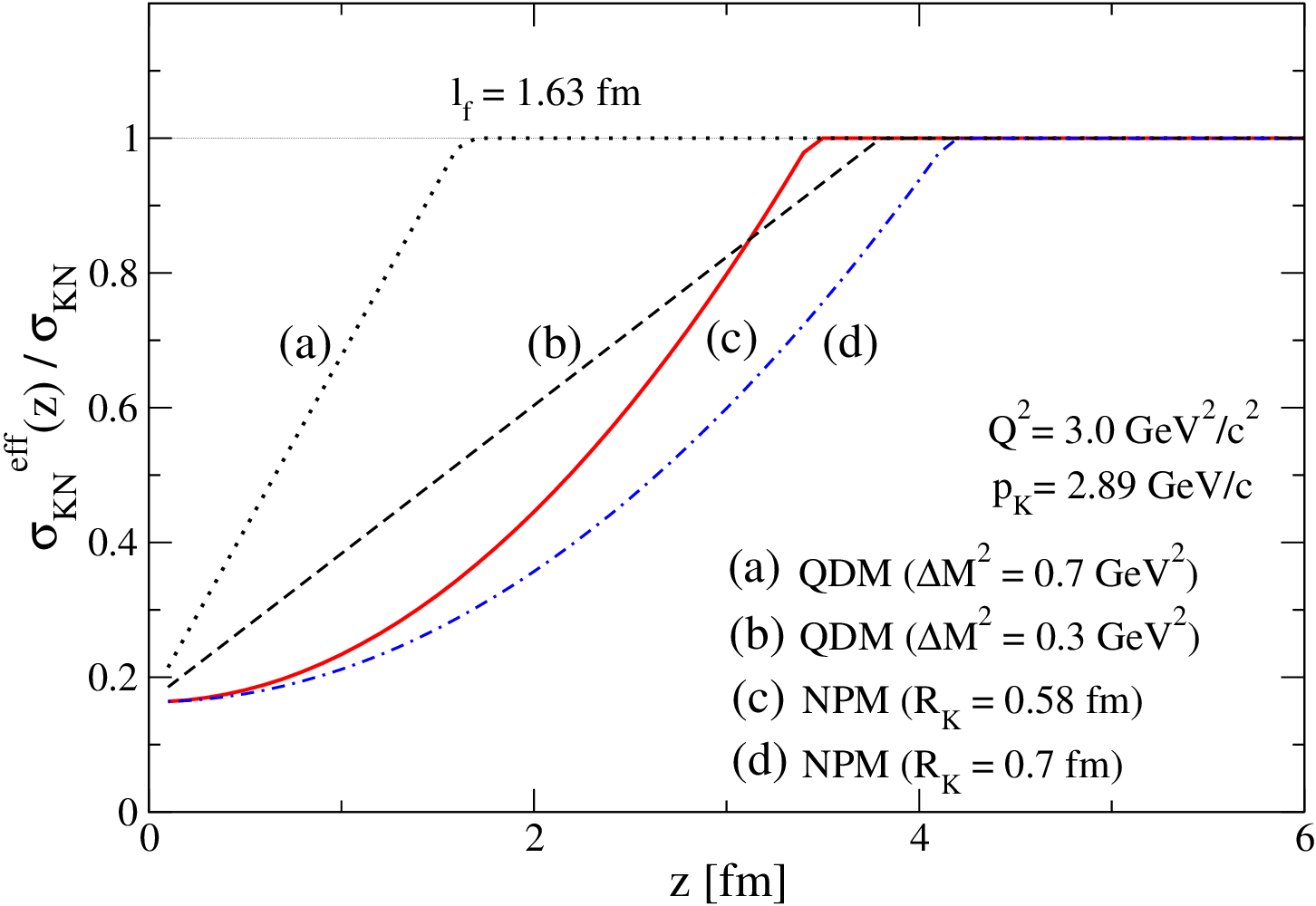}
\caption{ Ratio $\sigma_{KN}^{\rm eff}/\sigma_{KN}$ in
Eq.~(\ref{eq:sigmaeff}) as a function of the propagation distance
$z$ at $Q^2=3.0$ GeV$^2/c^2$ and $p_K=2.89$ GeV$/c$.
The four labeled curves correspond to the two
QDM cases, $l_f=1.63$ fm with $\Delta M^2=0.7$ GeV$^2$ and
$l_f=3.81$ fm with $\Delta M^2=0.3$ GeV$^2$, and the two NPM cases,
$l_f=3.45$ fm with $R_K=0.58$ fm and $l_f=4.16$ fm with $R_K=0.7$ fm.
The corresponding formation lengths are obtained from
Eqs.~(\ref{eq:lfnpm}) and (\ref{eq:lfqdm}) for the NPM and QDM,
respectively.
} \label{fig:fig1}
\end{figure}

For the NPM calculation we use $R_K=0.58$ fm in
Eq.~(\ref{eq:lfnpm}), corresponding to the measured kaon charge
radius,
$
\langle R_K^2\rangle = 0.34\pm 0.05~{\rm fm}^2,
$
reported in Ref.~\cite{amendolia}. For the QDM estimate we take $\Delta
M^2=0.3$ GeV$^2$, as advocated in Ref.~\cite{Das2019}, and compare it with
the more commonly used value $\Delta M^2=0.7$ GeV$^2$, which is
often favored in pion-transparency analyses.

Figure~\ref{fig:fig1} illustrates the ratio $\sigma_{KN}^{\rm
eff}/\sigma_{KN}$ in Eq.~(\ref{eq:sigmaeff}) as a function of the
propagation distance $z$ at $Q^2=3.0$ GeV$^2/c^2$ and $p_K=2.89$
GeV$/c$. The dotted, dashed, solid, and dash-dotted curves
correspond to the formation lengths $l_f=1.63$ fm with $\Delta
M^2=0.7$ GeV$^2$ and $l_f=3.81$ fm with $\Delta M^2=0.3$ GeV$^2$
from the QDM, and to $l_f=3.45$ fm with $R_K=0.58$ fm and
$l_f=4.16$ fm with $R_K=0.7$ fm from the NPM, respectively. The
figure makes the difference between the two scenarios transparent:
the QDM gives a linear rise with $z/l_f$, whereas the NPM yields a
quadratic increase. The contribution obtained with $\Delta
M^2=0.7$ GeV$^2$ is therefore noticeably weaker than the others.

\begin{figure}[t]
\centering
\includegraphics[width=0.95\linewidth]{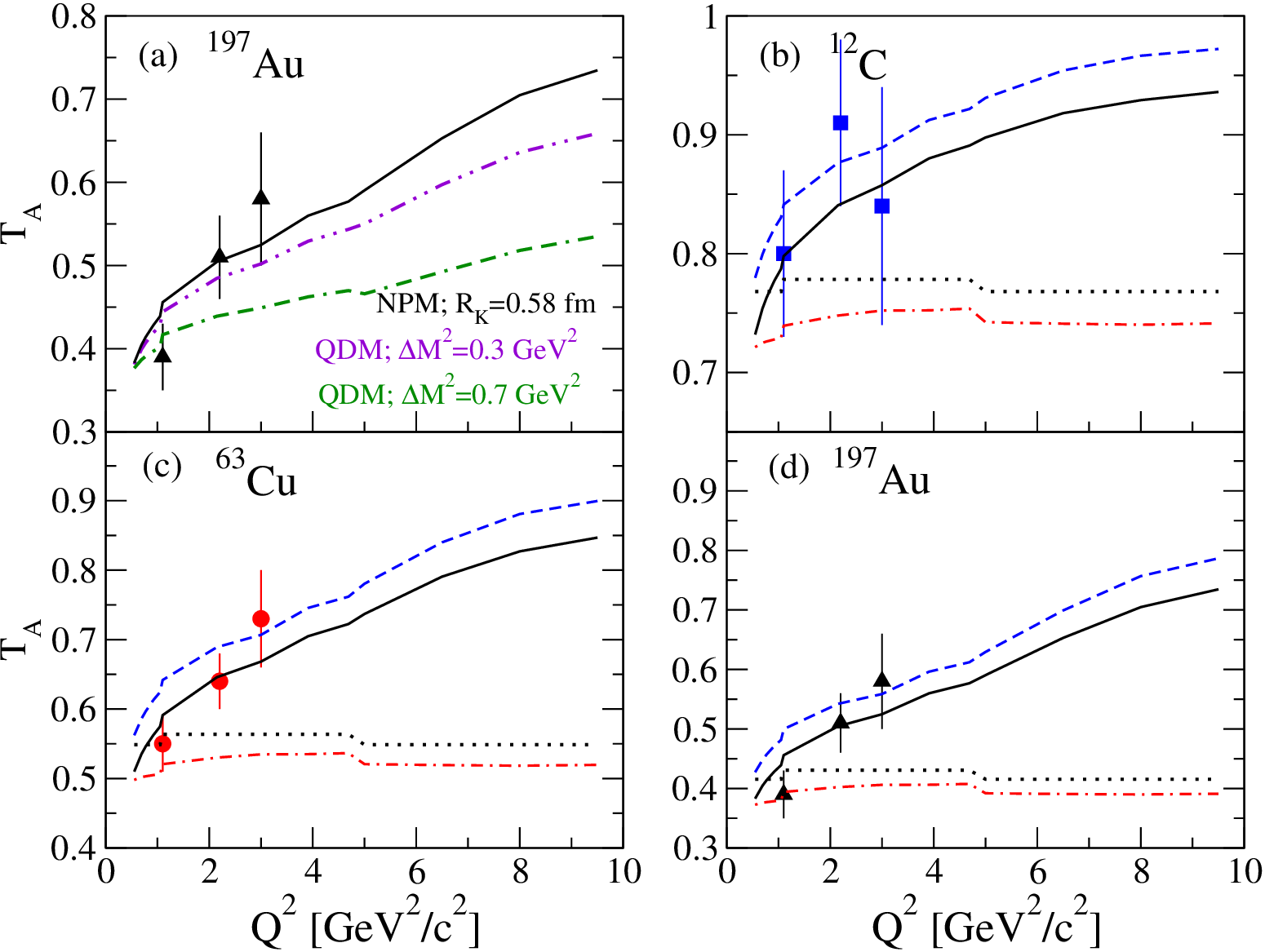}
\caption{
$Q^2$ dependence of the nuclear transparency $T_A$ in
$^{12}$C, $^{63}$Cu, and $^{197}$Au. The solid curves show the
result of the extended Glauber calculation in Eq.~(\ref{eq:TA}),
where the NPM with $R_K=0.58$ fm is used and the shadowing effect
is evaluated with $\sigma_{\phi N}=18$ mb. The dashed curves show
the NPM result without shadowing. The dotted curves correspond to
the pure Glauber calculation without NPM or shadowing, and the
dash-dotted curves indicate the shadowing correction to that
baseline. For comparison, panel (a) also shows the QDM predictions
for $\Delta M^2=0.3$ GeV$^2$ and $0.7$ GeV$^2$, represented by the
dash-dot-dotted and dash-dash-dotted curves, respectively.
The data are taken from Ref.~\cite{Nuruzzaman2011}.
}
\label{fig:fig2}
\end{figure}

The calculated $Q^2$ dependence of the transparency $T_A$ is shown in
Fig.~\ref{fig:fig2} for $^{12}$C, $^{63}$Cu, and $^{197}$Au. The solid
curves are obtained from the extended Glauber model of
Eq.~(\ref{eq:TA}), where the final-state CT effect is described by the
NPM with $R_K=0.58$ fm and the initial-state shadowing is evaluated
with $\sigma_{\phi N}=18$ mb. The dashed curves show the NPM result
without shadowing. The dotted curves correspond to the pure Glauber
baseline, where neither CT nor shadowing is included, and the
dash-dotted curves display the shadowing correction to that baseline.
For comparison, panel (a) also shows the QDM predictions with
$\Delta M^2=0.3$ and $0.7$~GeV$^2$ for the $^{197}$Au target.

Several features deserve emphasis. First, the inclusion of
shadowing lowers the transparency systematically and improves the
overall agreement with the data. Second, the QDM with the more
conventional choice $\Delta M^2=0.7$ GeV$^2$ fails to reproduce
the observed rise of the kaon transparency and is therefore not
retained in the subsequent comparison. Third, although the QDM
with $\Delta M^2=0.3$ GeV$^2$ becomes comparable to the NPM once
shadowing is included, such a choice is less well constrained
physically than the use of the experimental kaon charge radius in
the NPM. In addition, the steeper $Q^2$ dependence of the data is
more naturally reproduced by the quadratic growth of
$\sigma_{KN}^{\rm eff}(z;Q^2)$ in the NPM than by the linear rise
characteristic of the QDM.

\begin{figure}[t]
\centering
\includegraphics[width=0.95\linewidth]{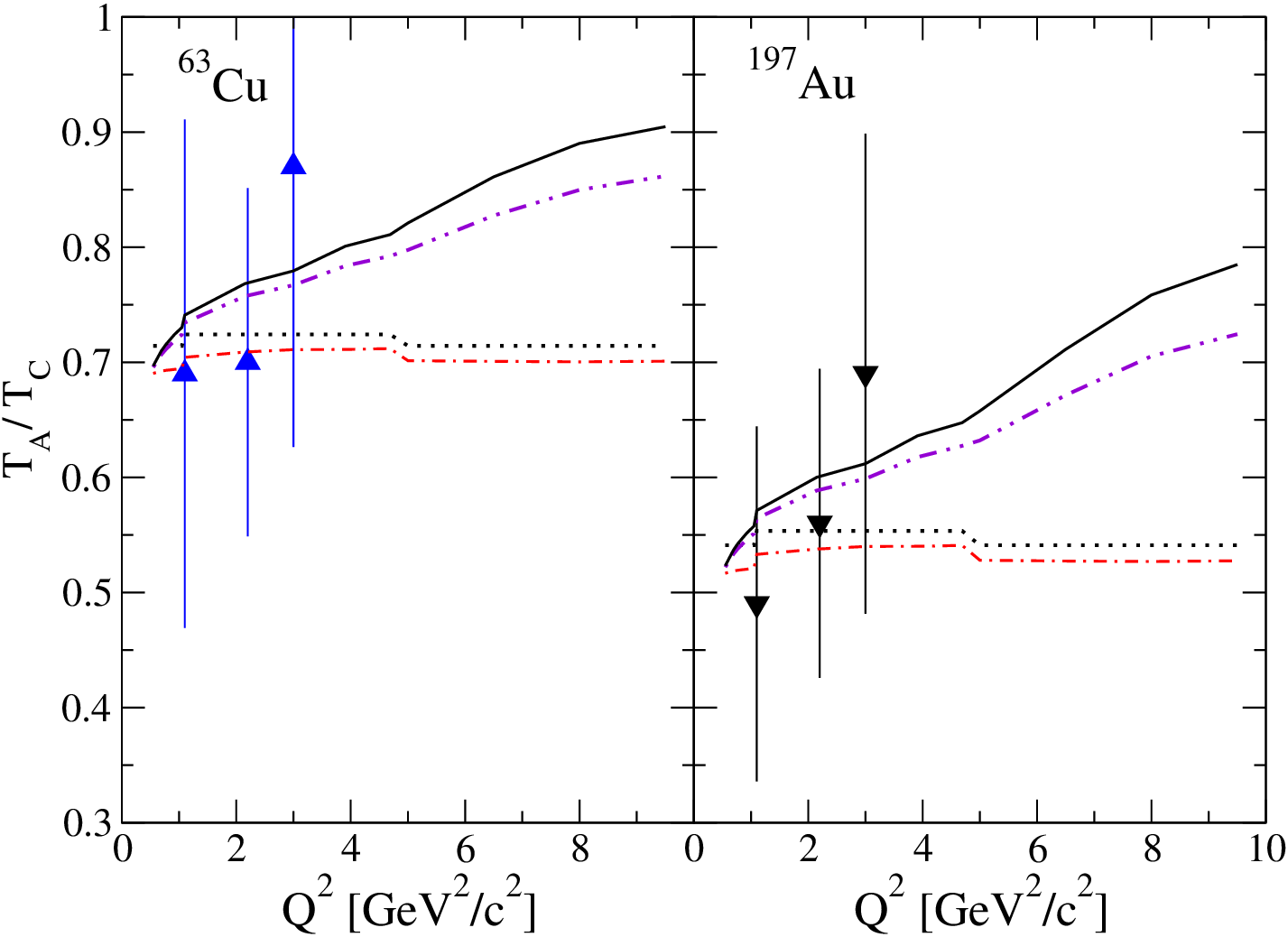}
\caption{
$Q^2$ dependence of the ratio $T_A/T_C$ obtained by normalizing the
transparency of each target nucleus to that of $^{12}$C. The solid
curves represent the ratio obtained from the extended Glauber
calculation including shadowing and the NPM, with the same notation as
in Fig.~\ref{fig:fig2}. The dash-dot-dotted curves show the
corresponding result from the extended Glauber calculation including
shadowing and the QDM with $\Delta M^2=0.3$ GeV$^2$. The data are taken
from Ref.~\cite{Nuruzzaman2011}.
} \label{fig:fig3}
\end{figure}

Figure~\ref{fig:fig3} presents the ratio $T_A/T_C$, where the
transparency of a heavier target is normalized to that of
$^{12}$C. Although normalization to deuterium is experimentally
advantageous, since the deuteron provides an approximately isoscalar
$pn$ system and can partially cancel effects from proton--neutron
imbalance, nuclear binding, and Fermi motion, such a normalization is
not entirely natural within the present framework. The Glauber model
used here is formulated for nuclei with sufficiently large $A$ and
nearly spherical density distributions, rather than for the weakly
bound and nonspherical deuteron. For this reason, we also consider the
ratio $T_A/T_C$, for which the normalization to $^{12}$C is better
matched to the conditions under which the present Glauber approach is
expected to be reliable. Unlike deuterium, however, $^{12}$C is not a
nearly transparent reference system, and its bound nucleons cannot be
regarded as quasi-free, which is a limitation of the $^{12}$C
normalization. Nevertheless, because $^{12}$C is already a finite
nucleus with a well-defined density distribution, one may incorporate
corrections associated with short-range correlations (SRC) and nuclear
attenuation, which can partly compensate for these effects. For this
reason, the $^{12}$C-normalized ratio remains a meaningful
supplementary observable within the present model. Within the present
framework, the shadowing-corrected NPM prediction (solid) exhibits a
steeper $Q^2$ dependence than the corresponding QDM prediction
(dash-dot-dotted) for both $^{63}$Cu and $^{197}$Au.

\begin{figure}[t]
\centering
\includegraphics[width=0.65\linewidth]{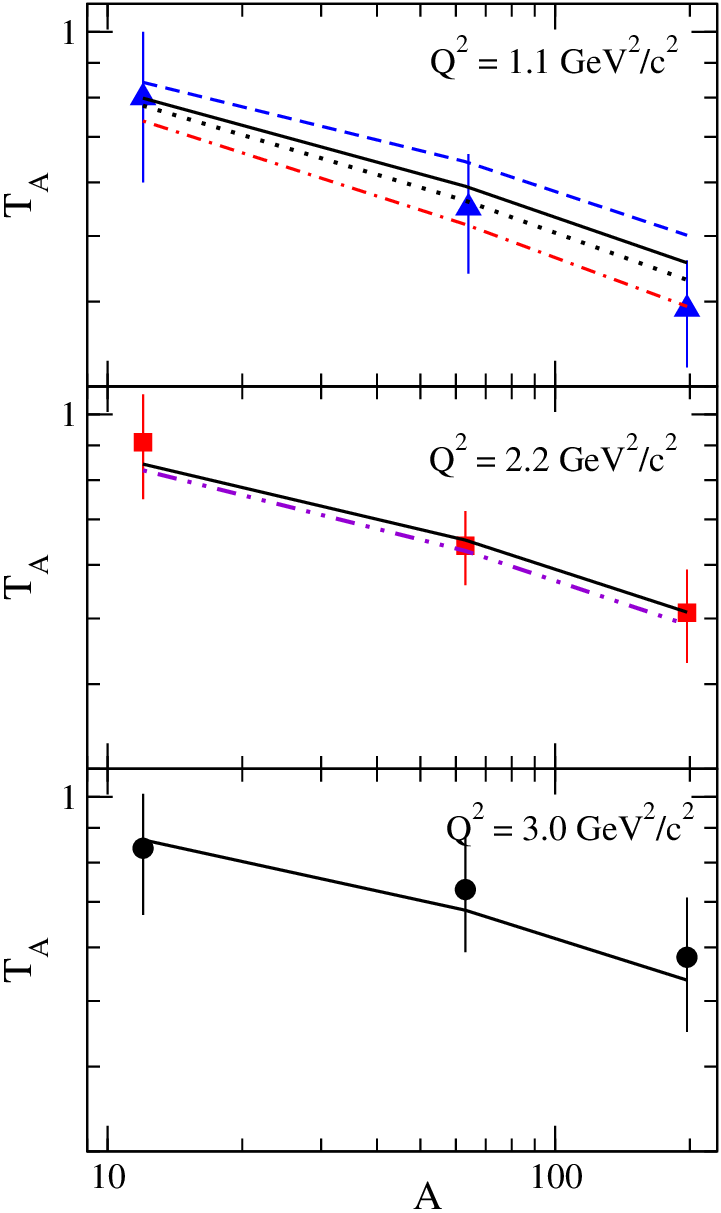}
\caption{
$A$ dependence of $T_A$ at fixed $Q^2$. From top to bottom, the panels
correspond to $Q^2=1.1$, 2.2, and 3.0 GeV$^2/c^2$, respectively. The
solid and dashed curves show the calculations with and without
shadowing, with the same notation as in Fig.~\ref{fig:fig2}. For
comparison, the QDM prediction at $Q^2=2.2$ GeV$^2/c^2$ is shown by the
dash-dot-dotted curve in the middle panel. The data are taken from
Ref.~\cite{Nuruzzaman2011}.
} \label{fig:fig4}
\end{figure}

The dependence of $T_A$ on the mass number $A$ is shown in
Fig.~\ref{fig:fig4} for fixed values of $Q^2$. From top to bottom, the
panels correspond to $Q^2=1.1$, 2.2, and 3.0 GeV$^2/c^2$, respectively.
As expected, the transparency decreases rapidly with increasing $A$,
reflecting the longer path of the outgoing kaon through nuclear matter
in heavier systems. The comparison between the solid and dashed curves
isolates the role of the initial-state shadowing, while the
dash-dot-dotted curve in the middle panel shows the QDM result for
comparison.

A related characterization of the nuclear dependence is provided
by the parameter $\alpha(Q^2)$, defined through
\begin{equation}
T_A=A^{\alpha(Q^2)-1},
\label{eq:alpha}
\end{equation}
which follows from the scaling form $\sigma_A=A^\alpha \sigma_N$.
A deviation of $\alpha$ from unity therefore signals a nuclear
effect beyond a simple incoherent sum over free nucleons. The
extraction of $\alpha(Q^2)$ has previously been discussed for pion
transparency~\cite{Qian2010} and for the kaon case in Ref.~\cite{Nuruzzaman2011}.

\begin{figure}[t]
\centering
\includegraphics[width=0.95\linewidth]{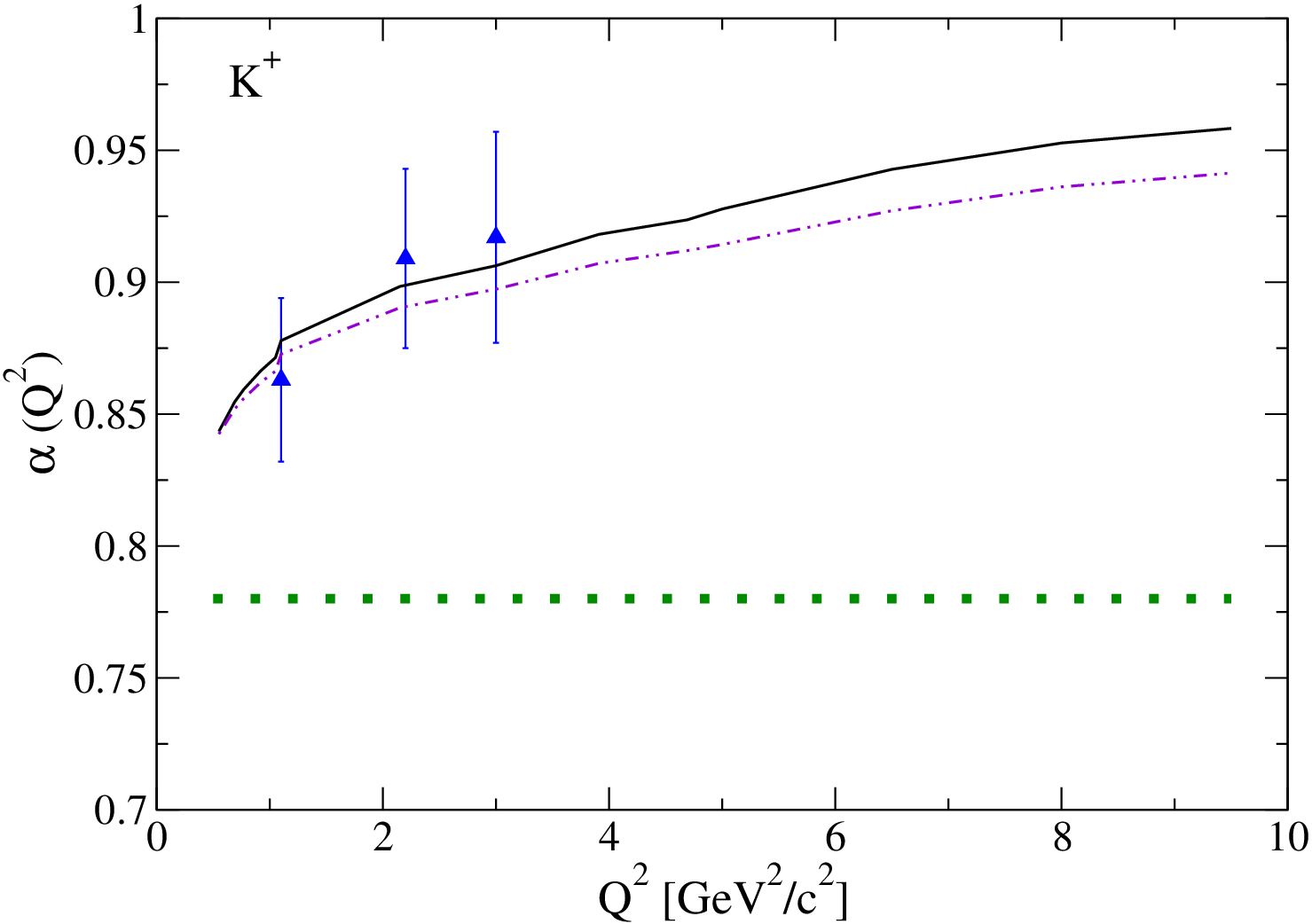}
\caption{
$Q^2$ dependence of $\alpha$ for the nuclear transparency $T_A$.
The parameter $\alpha$ is extracted by averaging over the JLab data
for the three target nuclei. The dash-dot-dotted curve shows the QDM
result for comparison, and the thick dotted line denotes the
approximately constant value of $\alpha$ extracted from
$K^+$--nucleus scattering~\cite{carroll}. The data are taken from
Ref.~\cite{Nuruzzaman2011}.
} \label{fig:fig5}
\end{figure}

The result is shown in Fig.~\ref{fig:fig5}. In hadron--nucleus
scattering, the empirical value $\alpha \simeq 0.78$ extracted from
$K^+$--nucleus data corresponds to an ordinary absorptive regime,
well below the volume-scaling limit $\alpha=1$, where
$T_A$ is independent of $A$, and still above the opaque-nucleus limit
$\alpha=2/3$ discussed by Carroll \textit{et al.}~\cite{carroll}.
By contrast, the rise of $\alpha(Q^2)$ from about 0.85 to 0.95 in
$A(e,e'K^+)$ indicates a substantial reduction of attenuation, so
that the transparency shows a progressively weaker $A$ dependence
and approaches more volume-like propagation in the nucleus.

\section{Conclusion}

We have investigated nuclear transparency in the reaction
$A(e,e'K^+)$ for the representative nuclei $^{12}$C, $^{63}$Cu, and
$^{197}$Au in order to analyze the onset of kaon color transparency.
Within an extended Glauber framework over the range
$0.55 \le Q^2 \le 10$ GeV$^2/c^2$, we included both the shadowing
associated with the two-step process in the initial state and the
color-transparency effect in the final state, described by either the
QDM or the NPM, and compared the results with the available JLab data
obtained with the 6-GeV electron beam.

Compared with the pion case under similar kinematical conditions,
strangeness production involves a smaller hadronic attenuation because
the relevant $K^+N$ cross section is weaker than the corresponding
$\pi N$ one over the momentum range of interest. In this situation,
the standard QDM parametrization constrained by the pion sector does
not account satisfactorily for the kaon data unless the mass parameter
is reduced to $\Delta M^2=0.3$ GeV$^2$. By contrast, the NPM, implemented
through the quadratic growth of $\sigma_{KN}^{\rm eff}(z;Q^2)$ and
supplemented by the initial-state shadowing correction, reproduces the
observed $Q^2$ dependence of $T_A$, the target-normalized ratio
$T_A/T_C$, and the overall $A$ dependence rather well. The present
analysis therefore indicates that the existing kaon-transparency data
favor the NPM-type expansion scenario over the standard QDM
parametrization, and that the steeper $Q^2$ dependence of kaon
transparency is more naturally accommodated by the NPM.

A more microscopic treatment within a dynamical transport framework
would nevertheless be valuable. To our knowledge, a dedicated study of
exclusive kaon transparency in $A(e,e'K^+)$ within a full GiBUU-type
or related dynamical transport framework has not yet been established
at the same level. In such approaches, effects such as Fermi motion,
multistep collisions, channel coupling, and collision broadening may
modify the quantitative attenuation, but are unlikely by themselves to
generate the steep monotonic $Q^2$ dependence observed in the present
data. We therefore expect that a full dynamical transport calculation
may change the quantitative values, but that the main qualitative
conclusion of the present work is likely to remain robust. The present
results also show that initial-state shadowing is not negligible even
in this energy regime and should be included together with final-state
color transparency when discussing the onset of CT in strangeness
production off nuclei.

%%%%%%%%%%%%%%%%%%%%%%%%%%%%%%%%%%%%%%%%%%%%%%%%%%%%%%%%%%%%%%%%%
       \section*{ACKNOWLEDGMENT}
%%%%%%%%%%%%%%%%%%%%%%%%%%%%%%%%%%%%%%%%%%%%%%%%%%%%%%%%%%%%%%%%%

This work was supported by the Grant No. NRF-2022R1A2B5B01002307
of the National Research Foundation (NRF) of Korea, and by the
Institute for Basic Science (IBS-R031-D1).

\end{document}